%% file: main.tex
\documentclass[sigconf]{acmart}

\usepackage[false]{anonymous-acm}

\AtBeginDocument{%
  }

\copyrightyear{2024}
\acmYear{2024}
\setcopyright{acmlicensed}\acmConference[ICDCN '24]{25th International Conference on Distributed Computing and Networking}{January 4--7, 2024}{Chennai, India}
\acmBooktitle{25th International Conference on Distributed Computing and Networking (ICDCN '24), January 4--7, 2024, Chennai, India}
\acmPrice{15.00}
\acmDOI{10.1145/3631461.3631559}
\acmISBN{979-8-4007-1673-7/24/01}

\acmConference[ICDCN '24]{25th International Conference on Distributed Computing and Networking}{January 4--7, 2024}{Chennai, India}

\begin{document}
\pdfoutput=1
\title{Performance evaluation of Private and Public Blockchains for multi-cloud service federation}

\authoranon{
    \author{Adam Zahir}
    \affiliation{%
      \institution{Universidad Carlos III de Madrid}
      \city{Madrid}
      \country{Spain}}
    \email{azahir@pa.uc3m.es}
    
    \author{Milan Groshev}
    \affiliation{%
      \institution{Universidad Carlos III de Madrid}
      \city{Madrid}
      \country{Spain}}
    \email{mgroshev@pa.uc3m.es}
    
    \author{Kiril Antevski}
    \affiliation{%
      \institution{AXo solutions}
      \city{Skopje}
      \country{Macedonia}}
    \email{kiril.antevski@axosolutions.com}
    
    \author{Carlos J.Bernardos}
    \affiliation{%
      \institution{Universidad Carlos III de Madrid}
      \city{Madrid}
      \country{Spain}}
    \email{cjbc@it.uc3m.es}
    
    \author{Constantine Ayimba}
    \affiliation{%
      \institution{Universidad Carlos III de Madrid}
      \city{Madrid}
      \country{Spain}}
    \email{ayconsta@it.uc3m.es}
    
    \author{Antonio de la Oliva}
    \affiliation{%
      \institution{Universidad Carlos III de Madrid}
      \city{Madrid}
      \country{Spain}}
    \email{aoliva@it.uc3m.es}
}
\renewcommand{\shortauthors}{A. Zahir et al.}
\acmArticleType{Research}
\acmCodeLink{https://github.com/borisveytsman/acmart}
\acmDataLink{htps://zenodo.org/link}

\begin{abstract}
The stringent low-latency, high reliability, availability and resilience requirements of 6G use cases will present challenges to cloud providers. Currently, cloud providers lack simple, efficient, and secure implementation of provisioning solutions that meet these challenges. Multi-cloud federation is a promising approach. In this paper, we evaluate the application of private and public blockchain networks for multi-cloud federation. We compare the performance of blockchain-based federation in private and public blockchain networks and their integration with a production-ready orchestration solution. Our results show that the public blockchain needs approximately 91 seconds to complete the federation procedure compared to the 48 seconds in the private blockchain scenario.
\end{abstract}

\maketitle
\input{introduction}
\input{related-work}
\input{DLT-federation}
\input{private-vs-public}
\input{experimental}

\input{conclusion}

\begin{acks}
This work has been partly funded by the European Commission Horizon Europe SNS JU DESIRE6G project, under grant agreement No.101096466, and the Spanish Ministry of Economic Affairs and Digital Transformation and the European Union-NextGenerationEU through the UNICO 5G I+D 6G-EDGEDT and 6G-DATADRIVEN.
\end{acks}

\balance
\bibliographystyle{ACM-Reference-Format}
\bibliography{acmart}
\balance
\clearpage

\end{document}

%% file: introduction.tex
\section{Introduction}
\label{introduction}
Edge computing has extended the cloud computing concept by creating a decentralized infrastructure in which computing resources are placed at the network edge. To support niche use cases (e.g., industrial automation, holographic telepresence, eHealth, gaming) public cloud providers (Public Clouds) host services at the network edge\cite{aws_green,Google_IoT}. Similarly, Telco operators and vertical industries (Private Clouds) use their private infrastructure to deploy Virtual Network Functions (VNFs) and enterprise workloads respectively. 

Given strict requirements on availability and resilience of 6G use cases \cite{6g-use-cases}, it is quite challenging for a lone computing network to fulfill such stringent Service-Level Agreements (SLAs). To improve elasticity, connectivity, or capacity, a cloud provider may need to leverage resources from different providers. This concept is known as multi-cloud federation~\cite{multi-cloud}. It enables service providers to; $(i)$ deal with peaks in service and resource requests, $(ii)$ replicate critical services, and $(iii)$ create backups to deal with disasters or scheduled inactivity~\cite{multi-cloud}. The potential benefits notwithstanding, Telco and IT cloud providers have reasonable doubts regarding the complexity, security, and practicality of implementing multi-cloud federation. The primary focus of service federation is to reduce execution time and enable the establishment of comprehensive end-to-end services across multiple providers. Achieving this goal requires careful negotiation and collaboration among the participating service providers.

A peer-to-peer individual approach offers security through trust distribution but is inefficient and slow in establishing relationships among a large number of unknown providers. Conversely, a centralized approach is fast and secure but requires a costly central entity that introduces a single point of failure.

A distributed solution can achieve a balance between decentralized and centralized approaches. Distributed Ledger Technologies (DLTs) such as Blockchain provide a secure, robust platform for both negotiation and execution of multi-cloud federation. Blockchain networks (private or public) are apt for this purpose due to their efficiency and interoperability. In particular, while efficient private blockchains are suitable for multi-cloud federation between different service providers~\cite{dlt-federation-1}, public blockchains can enable seamless user application federation where interaction between user applications and cloud providers can be triggered with a single on-chain transaction integrated into the user application. This distinctive capability of public blockchains can be of interest for a specific set of use cases (e.g., gaming, Virtual Reality) where the federation process can be triggered directly from the user application instead of asking the cloud provider to perform all the federation steps. Whereas blockchains are categorized based on the user access to verified/“trusted” (Private) and open access/"trustless" (Public), the cloud networks are categorized based on infrastructure access; leased (Public) or self-managed (Private). 

In this work, we investigate how public and private blockchain platforms affect the negotiation and execution of multi-cloud federation. Additionally, we analyze the performance of Blockchain-based federation on a production-ready Open Source Management and Orchestration (OSM) solution. To the best of our knowledge, this is the first study that conducts a performance analysis of multi-cloud federation using public and private blockchain platforms. 

The rest of the paper is organized as follows. In Section~\ref{sec:related} we position our work within the state-of-the-art. In Section~\ref{sec:dlt-federation} we describe the multi-cloud federation procedure and how blockchain can be applied. In Section~\ref{sec:prvate-vs-pub} we describe the private and public blockchains whose performance is evaluated later in Section~\ref{sec:experimental}. In Section~\ref{sec:conclusion} we present the conclusions and future directions of our work.

%% file: related-work.tex
\section{Related work}
\label{sec:related}
There is a growing interest in enabling the federation of services and resources in multi-cloud/multi-domain deployments. The goal is to improve the Quality-of-Service (QoS) / Quality-of-Experience (QoE) for both vertical customers and end users. 
Recent research projects (5G-Transformer~\cite{federation-2}, 5Growth~\cite{5growth}, and 5G-Coral~\cite{5gcoral}) have considered the orchestration of resources and services across multiple domains. Moreover, the SNS JU~\cite{sns-journal} research projects (Desire6G and Hexa-X-II), envision multi-cloud/multi-domain federation as an enabler of resilient cloud-native services.

Recent studies have contributed to the definition of network service federation~\cite{federation-1}, the architecture of the federation functionality~\cite{federation-2}, and the challenges and expectations of multi-cloud federation~\cite{multi-cloud}. In~\cite{decison-1}, the authors aim to maximize revenue for the administration domain, while ~\cite{decison-2} proposes a federation structure that maximizes social welfare. These approaches assume pre-existing SLAs among domains. In contrast, we propose a dynamic mechanism whereby service providers are allowed quick and brief federation relationships. 

In prior work, we proposed the application of blockchain to federation in dynamic environments~\cite{dlt-federation-1}, leveraging the technology to enable service providers reach short-term agreements in a secure and trustworthy manner. In~\cite{dlt-federation-2}, we apply blockchain federation to robotic services where mobile robots can extend their wireless range on demand. However, these studies are limited to the design and validation of blockchain technology and do not explicitly consider the nature of the blockchain itself. Conversely, in this work, we evaluate the performance of private and public blockchain platforms with regard to service federation. Moreover, we evaluate the performance of NVF Management and Orchestration (MANO) compatible orchestrator (OSM) for blockchain federation.  

%% file: DLT-federation.tex
\section{Blockchain based federation}
\label{sec:dlt-federation}

Network service federation is a feature of NFV MANO, that enables service providers to coordinate service deployment and migration across multiple administrative domains. This feature assists service providers in meeting the specific needs of vertical customers by reducing costs while making unused resources available to other services. Service providers can assume dual roles as both consumers and providers. Consumers request resources to deliver end-to-end network services to their own customers, while providers offer services and resources to consumers. 
In subsequent paragraphs, we discuss the federation process, how blockchain can enhance it and how blockchain federation fits in the NFV MANO architecture.

\subsection{Service federation procedures}
\label{subsec:federation-procedure}
There are several steps that network service providers need to follow in order to complete a service federation:

\begin{enumerate}
  \item \textbf{Registration}. Service providers register in a federation platform to interact with each other. Security mechanisms and validation procedures are essential to establish interactions.
  \item \textbf{Advertisement/Discovery}. Service providers periodically exchange or \textit{advertise} information regarding their services, according to the pre-established agreement. In cases where there is a large number of connections, a service provider may utilize polling or \textit{discovery} methods instead of direct advertising. By sharing and exchanging this information, a consumer service provider can gather a comprehensive overview of the available services within the federation.
  \item \textbf{Announcement/Negotiation}. 
  Federation begins when a consumer service provider broadcasts an announcement request for the federation. Potential providers can engage in a reverse-auction fashion by replying with bidding offers. The final decision is made by the consumer service provider using internal policy criteria. 
  \item \textbf{Deployment}. The “winning” provider proceeds with deploying the federated service. Upon successful deployment, both the consumer service provider and the provider of the federated service establish data plane connectivity and the inclusion of the federated service for the intended purpose.
  \item \textbf{Usage and Charging}. Once the federated service is embedded and running, the consumer service provider manages its life-cycle. Both service providers monitor the federated usage and calculate the fee according to the established agreement.
\end{enumerate}

Security, privacy, and trust are crucial in all the procedures outlined above. Furthermore, in the following section, we emphasize how blockchain as a technology preserves these key elements.

\subsection{Blockchain for federation}
\textit{Blockchain} technology originally emerged as a fundamental mechanism for Bitcoin, offering a distributed, secure, and timestamped ledger that records transactions among anonymous users. A blockchain consists of interconnected nodes sharing a single ledger that contains blocks of transactions of any data type, each with its own timestamp. Each block references the hash of the previous block, creating a chain of interconnected blocks that traces back to the genesis block (\textit{block 0}). The generation and validation of new blocks occur within a peer-to-peer blockchain network, where nodes (compute devices) participate. Blockchain networks can be public or private. To ensure the integrity of the blockchain, each new block must undergo validation through a consensus mechanism before being added to the blockchain. Various such mechanisms exist, offering different levels of security, trust, and privacy. In summary, applying blockchain technology in networking offers several advantages, such as enhanced security, data integrity, transparency, the elimination of the need for third-party intermediaries, and the integration of smart contracts \cite{advantages-blockchain-networking, advantages-blockchain-networking-2}.

Previous studies~\cite{dlt-federation-1} show that most of the federation challenges related to \textbf{Admission Control}, \textbf{Availability}, \textbf{Security and Privacy}, \textbf{Dynamic Pricing and Billing} and \textbf{Multi-domain QoS} can be addressed by deploying a non-customized (vanilla) version of a permissioned blockchain network (Ethereum, Hyperledger, Cosmos, Polkadot, etc.).

\subsection{Blockchain-based federation in NFV MANO}
Following the design proposed in~\cite{dlt-federation-1}, each cloud domain can be managed as an NFV MANO system, as depicted in Figure~\ref{fig:application_blockchain_NFVMANO}. NFV MANO is the ETSI standard architectural framework for managing and orchestrating virtualized and physical network functions and other software components. A blockchain node is installed within each service provider infrastructure with a direct connection to the Eastbound/Westbound interface of the NFV MANO framework. By adding and registering a node, the NFV MANO domains gain access to a private/public blockchain. 

Service requirements dictate which blockchain can be used. Private blockchains offer faster processing times and greater trust making them well-suited for closed federations where participants maintain control of the network. Conversely, public blockchains exhibit slower processing times and consume more computational resources given the larger number of nodes. They can be used by anyone for any purpose and are more robust to insider attacks, making them better suited for open and dynamic federations.

We compare the performance of public and private blockchains in the execution of service federation. In the following, we provide a brief description of both blockchains and proceed to discuss the results from our experimental testbed.

%% file: private-vs-public.tex
\section{Public and Private Blockchains}
\label{sec:prvate-vs-pub}
\begin{figure}[t]
    \vspace{-1em}
    \centering
    \includegraphics[width=0.8\columnwidth]{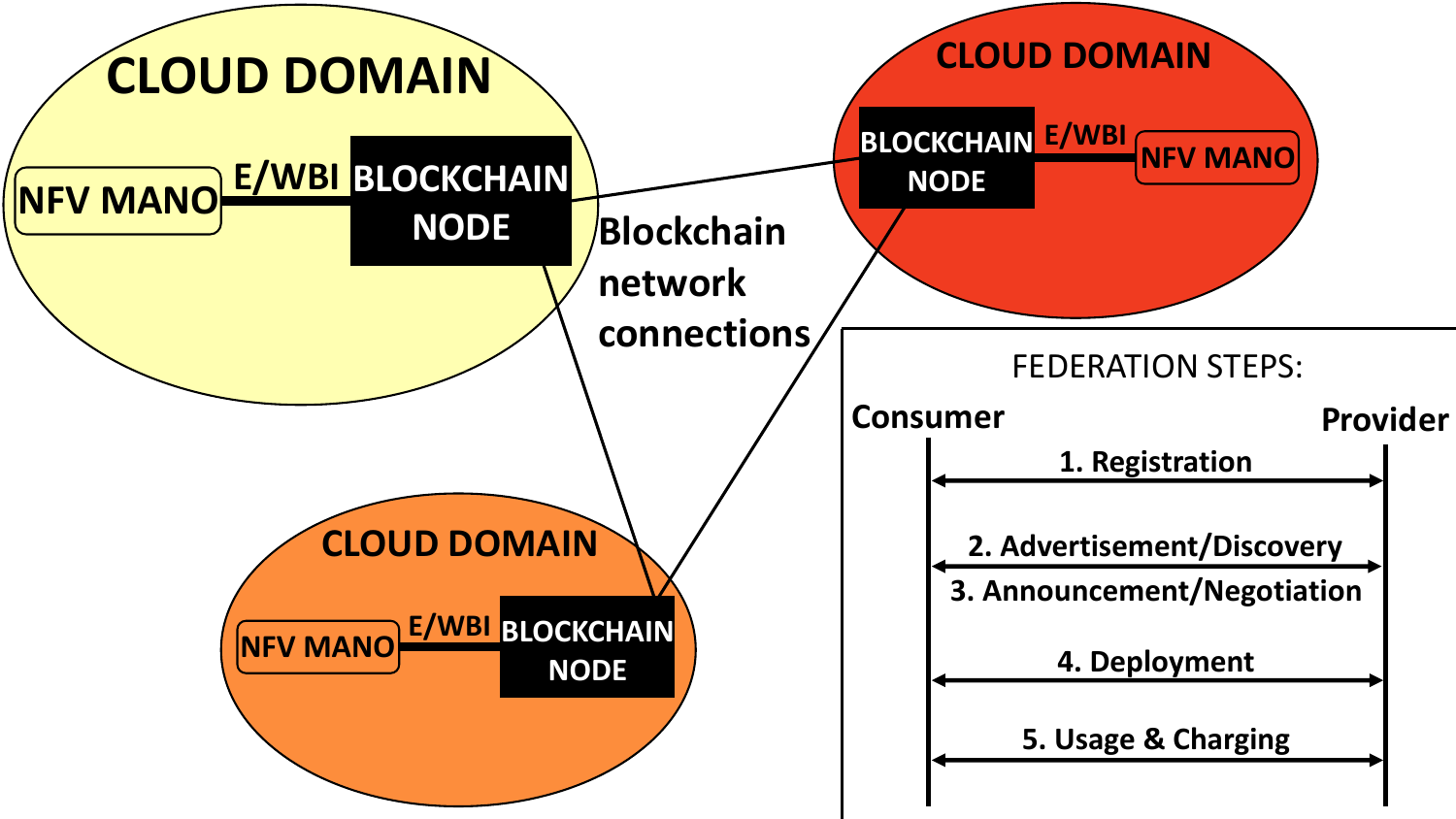}
    \caption{Blockchain application and steps for service federation in NFV MANO environments (adapted from ~\cite{dlt-federation-1})}
    \label{fig:application_blockchain_NFVMANO}
    \vspace{-1em}
\end{figure}
\subsection{Public Blockchain}
A public blockchain allows open participation. Any node can be admitted, validate transactions and execute the consensus mechanism. The decentralized nature prevents a single entity from controlling the network and reduces the risk of malicious activity from insiders. Typically, these networks are permissionless (no centralized administrator controlling participation) but others such as Alastria\footnote{\url{https://alastria.io/}} are permissioned: nodes join after verifying their identity and receiving specific permissions. A notable feature of public blockchains is transparency whereby on-chain information is publicly accessible enabling decentralized applications (smart contracts) to interact with each other via an open interface.

Smart contracts self execute when certain conditions are met. Consider an edge-cloud gaming scenario, if the gaming app detects a decrease in QoE and no nearby edge data centers are available, it can directly trigger the federation process on the blockchain without the intervention of the network operator. Thus, both smart contracts controlling the application and the federation interact within the same transaction.

However, this openness comes with a set of trade-offs, slow processing times due to network congestion, given the large number of nodes and high transaction volumes. Moreover, the secure consensus mechanism can lead to significant delays in transaction validation and confirmation compared to private blockchains. Additionally, the deployment and maintenance of smart contracts for applications can be expensive as a result of costly transaction fees. 

\subsection{Private Blockchain} 
A private blockchain is controlled by an organization that only admits verified members. This generates higher trust among participants. Furthermore, private blockchains process a significantly reduced volume of transactions, resulting in lower competition for block space. This enables the use of consensus algorithms with lower security requirements and faster processing times. However, the information stored remains private among the admitted parties, limiting on-chain interaction with other applications.

A similar context where private blockchains can be advantageous is during a football match with a sudden surge in users. By leveraging a blockchain and using a set of transactions, Telco operators can federate their radio resources with one another to meet unforeseen capacity demands.

The restricted access and control of the network by authorized entities concentrates the maintenance cost to the limited number of users. The trade-off of employing weaker consensus mechanisms for better performance may lead to data manipulation, transaction censorship, and centralized power by selected users. 

%% file: experimental.tex
\section{Federation performance of public and private blockchains}
\label{sec:experimental}
In this section, we first describe the hardware and software setup of our test bed then the data collection process. Lastly, we discuss the obtained results and our contribution to the state of the art in multi-cloud service federation. Given the current absence of open-source standardized solutions for federation, our evaluation focuses on blockchain-based approaches. However, in the future extension of this work, we plan to include comparisons with other federation approaches proposed by ongoing research projects.

\subsection{Experimental setup}
\label{sec:experimental_setup}
\begin{figure}[t]
    \centering
    \includegraphics[width=0.8\columnwidth]{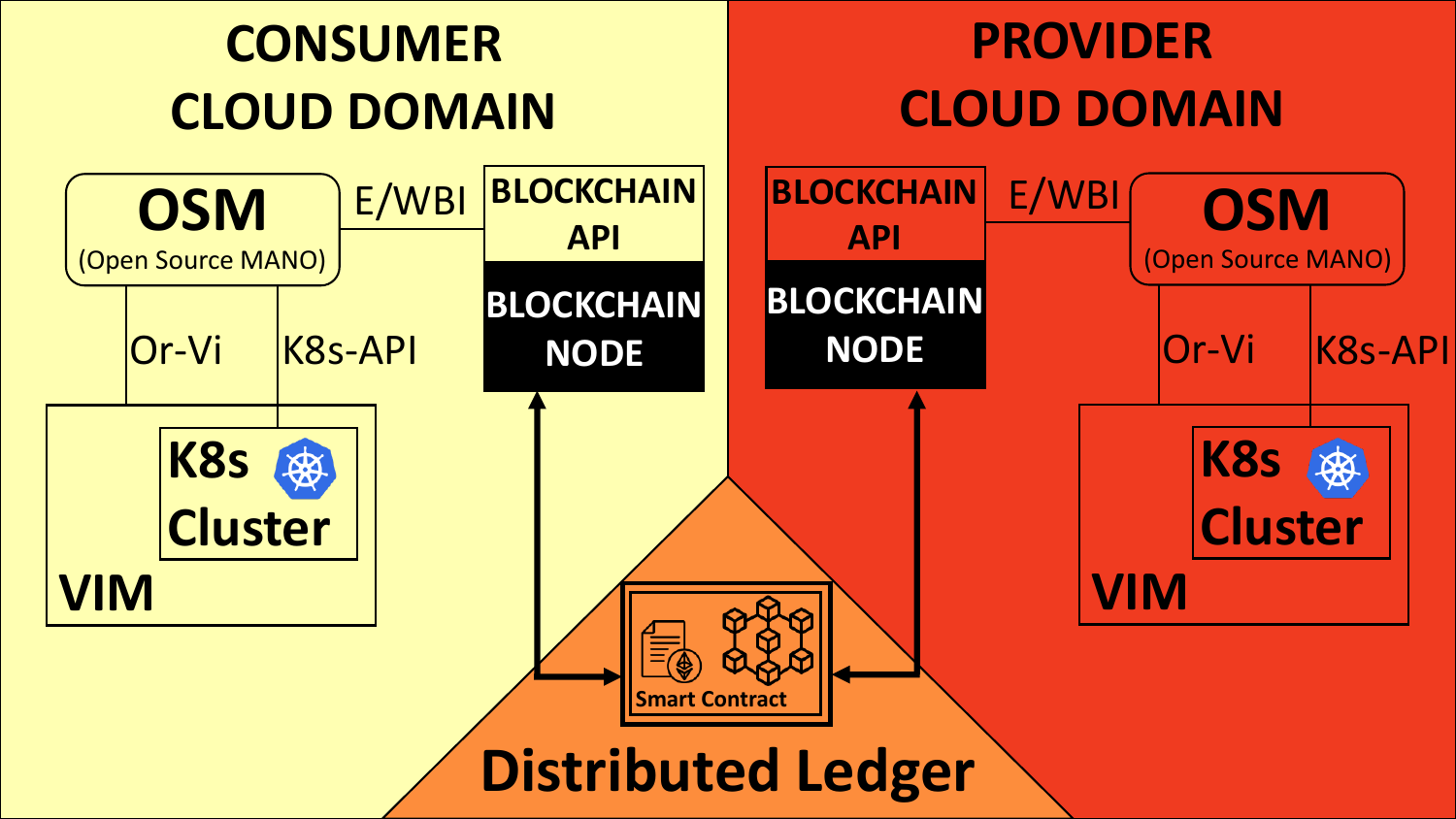}
    \caption{Experimental setup}
    \label{fig:experimental_setup}
    \vspace{-1em}
\end{figure}

Our experimental setup is shown in Figure \ref{fig:experimental_setup}, it includes two cloud domains -- consumer and provider -- and their underlying infrastructure, an orchestrator and a blockchain node. Each domain is deployed as a virtual machine with 4 CPU cores, 8 GB of RAM, and 80 GB of disk memory running Ubuntu Server 20.04.

\begin{figure}[t]
    \centering
    \includegraphics[width=0.8\columnwidth]{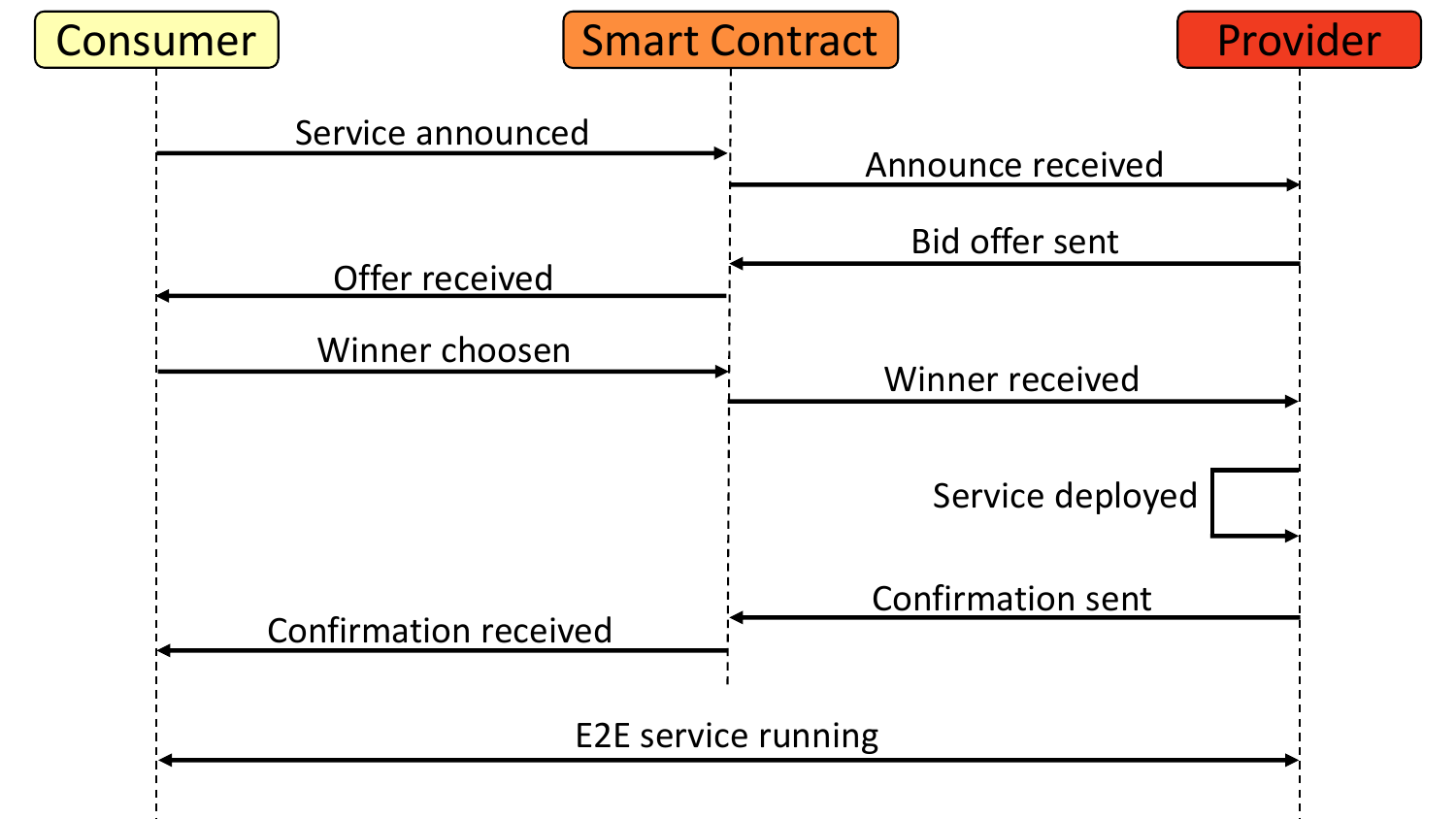}
    \caption{Sequence diagram of Experimental setup}
    \label{fig:sequence_diagram_experimental_setup}
    \vspace{-1em}
\end{figure}

In each domain, the orchestration platform, for managing the VNFs is Open Source MANO (OSM)\footnote{\url{https://osm.etsi.org/}}. The federation services used consist of Kubernetes\footnote{\url{https://kubernetes.io/}} Network Functions (KNFs), instantiated through indirect calls to the Kubernetes API with Helm Charts\footnote{\url{https://helm.sh/}}. Therefore, each domain has a dummy Virtualized Infrastructure Manager (VIM) account associated with a Kubernetes Cluster, both registered in OSM. All federation procedures are stored and deployed on a single smart contract running on the blockchain. Given the absence of an implemented E/WBI in OSM, we developed a Blockchain-based API to interact with the blockchain network and Federation smart contract. We employed the Ethereum platform~\cite{ethereum} with \textit{Proof-of-Authority} (PoA) consensus mechanism for both private and public blockchains. We configured a two-node private blockchain using Go-Ethereum (Geth)\footnote{\url{https://geth.ethereum.org/}}. For the public blockchain, we used the Goerli Testnet\footnote{\url{https://goerli.net/}} and accessed it via the Infura service, which provides a public Ethereum API. This approach allowed for a more efficient and cost-effective network access.

In our experiments, the consumer sends a transaction in the smart contract to announce the desired service deployment with specific requirements. The provider actively listens for federation events and responds by generating a bid-offer transaction with the service price. Subsequently, the consumer selects the offer from the provider domain and the winning provider deploys the requested federated service. We employed a simple Kubernetes service using an nginx image and a load balancer for connectivity between the deployed service and the consumer domain. After successful deployment, the provider notifies the consumer and shares relevant service information, including connection details. Figure \ref{fig:sequence_diagram_experimental_setup} illustrates the sequence of steps involved in the process.

\subsection{Experimental Results}

We carried out two sets of experiments as described in Section \ref{sec:experimental_setup} to measure the average time of each step in the federation process. We collected these data by recording the precise timestamps of transaction publications and event receptions in the client. Our first experiments involved a private blockchain comprising two nodes. We conducted several tests by altering the "period" parameter within the genesis block. This parameter determines the average time interval for adding new blocks to the blockchain, which may vary depending on factors such as the smart contract's complexity and the number of transactions included in each block. We tested block periods (BP) of [1s, 2s, 5s, 10s, 20s] to analyze the network performance under different conditions. As the network size increases, the BP has to increase to ensure efficient propagation of transactions and blocks throughout the network. In the second set of experiments, we replicated the procedure on the public blockchain to compare the results with the best and worst private block times obtained in the previous tests. Notably, the BP in the public blockchain is determined by the network size and is beyond our control.

\begin{figure}[t]
    \centering
    \includegraphics[width=0.8\columnwidth]{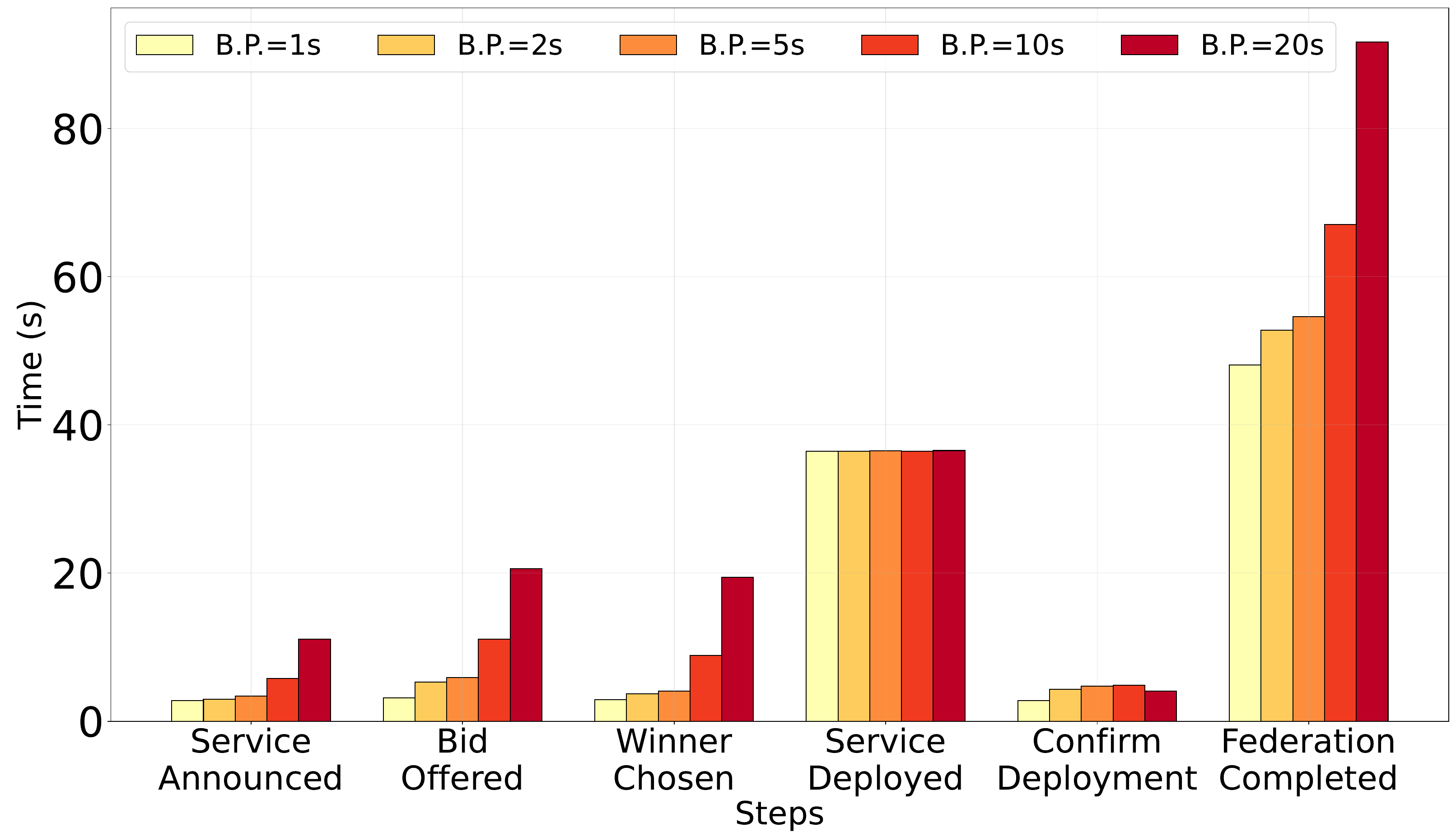}
    \caption{Federation process duration on a private blockchain}
    \label{fig:private_results}
    \vspace{-1em}
\end{figure}
\begin{figure}[t]
    \centering
    \includegraphics[width=0.8\columnwidth]{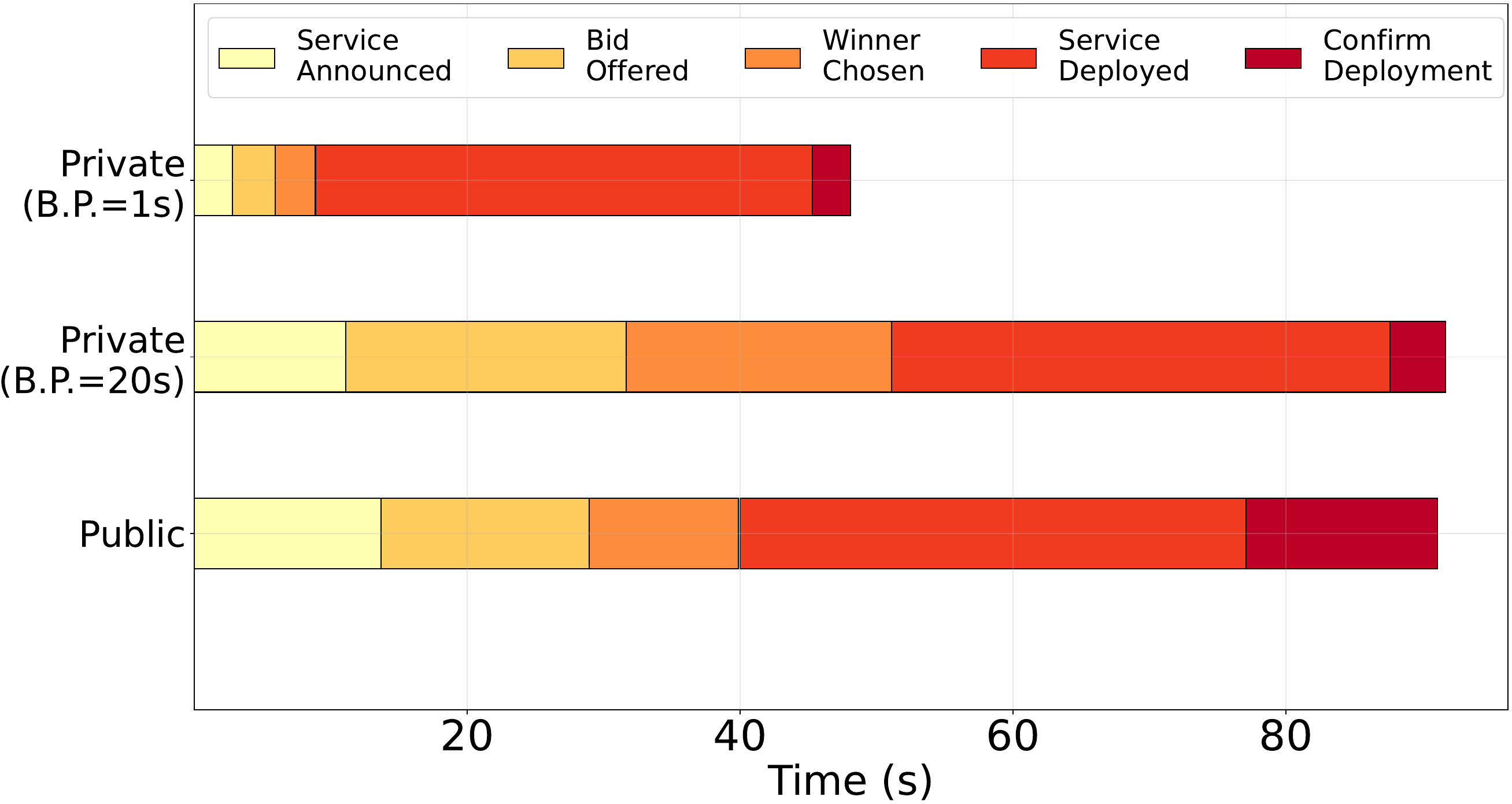}
    \caption{Comparison between the most and least favorable private block periods in relation to the public blockchain}
    \label{fig:private_vs_public_results}
    \vspace{-1em}
\end{figure}

Figure \ref{fig:private_results} shows the event plot of the first set of experiments, comprising 100 averaged measurements. On the y-axis is the duration in seconds of each federation step, while the x-axis represents the different BPs. Each BP is represented with a color scheme ranging from lighter shades for short BPs (1s) to darker shades for longer BPs (20s). From left to right, we can observe a similar pattern for the \textit{Service Announced}, \textit{Bid Offered}, and \textit{Winner Chosen} procedures. For the latter, increasing the BP up to 5s does not significantly change the federation times. However, when the BP is increased from 5s to 10s and from 10s to 20s, federation time nearly doubles. This is because, during the negotiation phases, various data are exchanged as part of the transactions. When the BP is extended, it takes longer to verify and add these transactions to the blockchain. Results indicate that the network service instantiation time in OSM (\textit{Service Deployed} procedure) dominates, taking approximately 36s to complete. This behavior is the result of the onboarding time for the network functions and network services descriptors, and the creation time of the Kubernetes Service in the cluster. 
In the penultimate phase, involving the confirmation of service deployment, the time duration is quite similar for all BPs. This is because this step involves lighter transactions and reduced reading operations (i.e., read data from blocks via API calls), which requires fewer computational resources. Finally, the \textit{Federation Completed} phase presents the accumulated times of all federation procedures.

Figure \ref{fig:private_vs_public_results} provides a comparison between the most and least favorable private BP in relation to the public blockchain. The graph represents the time on the x-axis and the type of blockchain network on the y-axis, with colors indicating the different federation procedures. We can notice that, when using a public blockchain, the average time to complete a federation is \textasciitilde 91 seconds, whereas the corresponding time using a private blockchain is \textasciitilde 48 seconds in the best case and \textasciitilde 92 seconds in the worst case. The behavior of the public network is comparable to the private network with a BP of 20s in certain phases, such as \textit{Service Announced} and \textit{Bid Offered}. However, we can observe significant time variations in the \textit{Winner Chosen} and \textit{Confirm Deployment} procedures. These differences are mainly due to the distinct sizes of the blockchains. The private blockchain, involving only 2 participants, demonstrated more consistent and efficient performance. In contrast, the public Goerli Testnet presented lower stability due to its highly condensed nature and the timing of experimentation. 

Results show that public blockchains can be advantageous in scenarios where multiple entities, such as cloud providers, private edge data centers, hybrid clouds, and network operators are involved, bearing in mind the potential delays caused by network congestion. However, when participants consist of known and trusted entities (e.g. only Telco operators), a private blockchain is more suitable and offers better performance. It is worth noting that security and scalability issues in blockchain technology are primarily related to the consensus protocol. In the context of multi-cloud federation, private blockchains can directly mitigate security issues by offering controlled user admission as well as operators that run the blockchain nodes. This observation aligns with the "Blockchain trilemma" concept coined by Ethereum's founder, Vitalik Buterin, which argues that a blockchain can either have scalability and decentralization but weak security, or high decentralization and security but weak scalability. Therefore, depending on the use case, we propose the use of different blockchains with apt consensus mechanisms to trade off properties in the blockchain trilemma.

Regarding the cost implications of implementing either solution, we refer the reader to Table 1 in~\cite{dlt-federation-3}, which provides a comprehensive comparison of multiple blockchain consensus algorithms across different platforms. While the table focuses on private blockchains, it also includes insights into the portability to public chains, exposing the associated costs and performance factors.

%% file: conclusion.tex
\section{Conclusion and Future Work}
\label{sec:conclusion}
In this paper, we have analyzed how private and public blockchain networks can be applied to the process of multi-cloud federation using a production-ready orchestration solution. We successfully set up an experimental scenario where we deployed an OSM orchestration solution that uses Kubernetes as a VIM to deploy a simple application service. We tested public and private Ethereum-based blockchains for the multi-cloud federation. Our results show that both blockchain platforms have their own benefits and drawbacks, and the choice mainly depends on the type of user application. 

As future work, we plan to optimize the OSM Blockchain-federation deployment time. For what concerns the Blockchain, we plan to expand the federation features for different Telco procedures, such as healing and migration. Additionally, we plan to integrate our "open-end innovation" approach within a real user application and study the direct triggering of the federation process with a single on-chain transaction. This integration will allow us to evaluate the performance of our approach within a real-world context, providing insights into its potential benefits and challenges.

%% file: main.bbl

\begin{thebibliography}{17}


\ifx \showCODEN    \undefined \def \showCODEN     #1{\unskip}     \fi
\ifx \showDOI      \undefined \def \showDOI       #1{#1}\fi
\ifx \showISBNx    \undefined \def \showISBNx     #1{\unskip}     \fi
\ifx \showISBNxiii \undefined \def \showISBNxiii  #1{\unskip}     \fi
\ifx \showISSN     \undefined \def \showISSN      #1{\unskip}     \fi
\ifx \showLCCN     \undefined \def \showLCCN      #1{\unskip}     \fi
\ifx \shownote     \undefined \def \shownote      #1{#1}          \fi
\ifx \showarticletitle \undefined \def \showarticletitle #1{#1}   \fi
\ifx \showURL      \undefined \def \showURL       {\relax}        \fi
\providecommand\bibfield[2]{#2}
\providecommand\bibinfo[2]{#2}
\providecommand\natexlab[1]{#1}
\providecommand\showeprint[2][]{arXiv:#2}

\bibitem[aws({[n.\,d.]})]%
        {aws_green}
 \bibinfo{year}{[n.\,d.]}\natexlab{}.
\newblock \bibinfo{title}{{AWS Greengrass}}.
\newblock \bibinfo{howpublished}{\url {https://aws.amazon.com/greengrass/}}.
\newblock
\newblock
\shownote{Accessed: 2023-09-21}.


\bibitem[Goo({[n.\,d.]})]%
        {Google_IoT}
 \bibinfo{year}{[n.\,d.]}\natexlab{}.
\newblock \bibinfo{title}{{Google IoT Core}}.
\newblock \bibinfo{howpublished}{\url {https://cloud.google.com/iot-core}}.
\newblock
\newblock
\shownote{Accessed: 2023-09-21}.


\bibitem[{6G SNS}(2023)]%
        {sns-journal}
\bibfield{author}{\bibinfo{person}{{6G SNS}}.} \bibinfo{year}{2023}\natexlab{}.
\newblock \bibinfo{title}{{SNS Journal/2023}}.
\newblock
\newblock


\bibitem[Antevski and Bernardos(2022)]%
        {dlt-federation-1}
\bibfield{author}{\bibinfo{person}{Kiril Antevski} {and} \bibinfo{person}{Carlos~J. Bernardos}.} \bibinfo{year}{2022}\natexlab{}.
\newblock \showarticletitle{Federation in Dynamic Environments: Can Blockchain Be the Solution?}
\newblock \bibinfo{journal}{\emph{IEEE Communications Magazine}} (\bibinfo{year}{2022}).
\newblock


\bibitem[Antevski and Bernardos(2023)]%
        {dlt-federation-3}
\bibfield{author}{\bibinfo{person}{Kiril Antevski} {and} \bibinfo{person}{Carlos~J Bernardos}.} \bibinfo{year}{2023}\natexlab{}.
\newblock \showarticletitle{Applying Blockchain consensus mechanisms to Network Service Federation: Analysis and performance evaluation}.
\newblock \bibinfo{journal}{\emph{Computer Networks}} (\bibinfo{year}{2023}).
\newblock


\bibitem[et~al.(2020a)]%
        {dlt-federation-2}
\bibfield{author}{\bibinfo{person}{Antevski et al.}} \bibinfo{year}{2020}\natexlab{a}.
\newblock \showarticletitle{DLT federation for Edge robotics}. In \bibinfo{booktitle}{\emph{IEEE NFV-SDN 2020}}.
\newblock


\bibitem[et~al.(2020b)]%
        {decison-1}
\bibfield{author}{\bibinfo{person}{Antevski et al.}} \bibinfo{year}{2020}\natexlab{b}.
\newblock \showarticletitle{A Q-learning strategy for federation of 5G services}. In \bibinfo{booktitle}{\emph{IEEE ICC 2020}}.
\newblock


\bibitem[et~al.(2020c)]%
        {federation-2}
\bibfield{author}{\bibinfo{person}{Baranda~Hortiguela et al.}} \bibinfo{year}{2020}\natexlab{c}.
\newblock \showarticletitle{5G-TRANSFORMER meets Network Service Federation: design, implementation and evaluation}. In \bibinfo{booktitle}{\emph{IEEE NetSoft 2020}}.
\newblock


\bibitem[et~al.(2020d)]%
        {6g-use-cases}
\bibfield{author}{\bibinfo{person}{Giordani et al.}} \bibinfo{year}{2020}\natexlab{d}.
\newblock \showarticletitle{Toward 6G Networks: Use Cases and Technologies}.
\newblock \bibinfo{journal}{\emph{IEEE Communications Magazine}} (\bibinfo{year}{2020}).
\newblock


\bibitem[et~al.(2021)]%
        {5growth}
\bibfield{author}{\bibinfo{person}{Guimarães et al.}} \bibinfo{year}{2021}\natexlab{}.
\newblock \showarticletitle{Public and Non-Public Network Integration for 5Growth Industry 4.0 Use Cases}.
\newblock \bibinfo{journal}{\emph{IEEE Communications Magazine}} (\bibinfo{year}{2021}).
\newblock


\bibitem[et~al.(2018a)]%
        {federation-1}
\bibfield{author}{\bibinfo{person}{Li et al.}} \bibinfo{year}{2018}\natexlab{a}.
\newblock \showarticletitle{Service orchestration and federation for verticals}. In \bibinfo{booktitle}{\emph{IEEE WCNCW 2018}}.
\newblock


\bibitem[et~al.(2018b)]%
        {5gcoral}
\bibfield{author}{\bibinfo{person}{Rapone et al.}} \bibinfo{year}{2018}\natexlab{b}.
\newblock \showarticletitle{An Integrated, Virtualized Joint Edge and Fog Computing System with Multi-RAT Convergence}. In \bibinfo{booktitle}{\emph{IEEE BMSB 2018}}.
\newblock


\bibitem[ETSI(2020)]%
        {advantages-blockchain-networking}
\bibfield{author}{\bibinfo{person}{ETSI}.} \bibinfo{year}{2020}\natexlab{}.
\newblock \showarticletitle{ETSI ISG PDL 003 V1.1.1, Permissioned Distributed Ledger (PDL); Application Scenarios}.
\newblock


\bibitem[ETSI(2021)]%
        {advantages-blockchain-networking-2}
\bibfield{author}{\bibinfo{person}{ETSI}.} \bibinfo{year}{2021}\natexlab{}.
\newblock \showarticletitle{ETSI ISG PDL 003 V1.1.1, Permissioned Distributed Ledger (PDL); Smart Contracts System Architecture and Functional Specification}.
\newblock


\bibitem[Petcu(2013)]%
        {multi-cloud}
\bibfield{author}{\bibinfo{person}{Dana Petcu}.} \bibinfo{year}{2013}\natexlab{}.
\newblock \showarticletitle{Multi-Cloud: Expectations and Current Approaches}. In \bibinfo{booktitle}{\emph{Proceedings of the 2013 International Workshop on Multi-Cloud Applications and Federated Clouds}}.
\newblock


\bibitem[Tai and Yen(2021)]%
        {decison-2}
\bibfield{author}{\bibinfo{person}{Yu-Chen Tai} {and} \bibinfo{person}{Li-Hsing Yen}.} \bibinfo{year}{2021}\natexlab{}.
\newblock \showarticletitle{Network Service Embedding in Multiple Edge Systems: Profit Maximization by Federation}. In \bibinfo{booktitle}{\emph{IEEE ICC 2021}}.
\newblock


\bibitem[Wood et~al\mbox{.}(2014)]%
        {ethereum}
\bibfield{author}{\bibinfo{person}{Gavin Wood} {et~al\mbox{.}}} \bibinfo{year}{2014}\natexlab{}.
\newblock \showarticletitle{Ethereum: A secure decentralised generalised transaction ledger}.
\newblock \bibinfo{journal}{\emph{Ethereum project yellow paper}} (\bibinfo{year}{2014}).
\newblock


\end{thebibliography}
